\documentclass[
superscriptaddress,
amsmath,
amssymb,
aps,
nolongbibliography
pra,
floatfix,
twocolumn
]{revtex4-2}
\usepackage{url}
\usepackage[
]{hyperref}
\usepackage{graphicx}
\usepackage{dcolumn}
\usepackage{bm}

\begin{document}

	\title{Non-Abelian braiding on photonic chips}
	
	\author{Xu-Lin Zhang}
	\affiliation{State Key Laboratory of Integrated Optoelectronics, College of Electronic Science and Engineering,\\
	Jilin University, Changchun 130012, China}
	\author{Feng Yu}
	\affiliation{State Key Laboratory of Integrated Optoelectronics, College of Electronic Science and Engineering,\\
	Jilin University, Changchun 130012, China}
	\author{Ze-Guo Chen}
	\affiliation{Department of Physics, Hong Kong Baptist University, Kowloon Tong, Hong Kong, China}
	\author{Zhen-Nan Tian}
	\affiliation{State Key Laboratory of Integrated Optoelectronics, College of Electronic Science and Engineering,\\
	Jilin University, Changchun 130012, China}
	\author{Hong-Bo Sun}
	\affiliation{State Key Laboratory of Integrated Optoelectronics, College of Electronic Science and Engineering,\\
	Jilin University, Changchun 130012, China}
	\affiliation{State Key Laboratory of Precision Measurement Technology and Instruments, Department of Precision Instrument,\\
	Tsinghua University, Haidian, Beijing 100084, China}
	\author{Guancong Ma}
	\affiliation{Department of Physics, Hong Kong Baptist University, Kowloon Tong, Hong Kong, China}
	\date{March 29, 2022}
	\begin{abstract}
		Non-Abelian braiding has attracted significant attention because of its pivotal role in describing the exchange behaviors of anyons–a candidate for realizing quantum logics. The input and outcome of non-Abelian braiding are connected by a unitary matrix which can also physically emerge as a geometric-phase matrix in classical systems. Hence it is predicted that non-Abelian braiding should have analogues in photonics, but a feasible platform and the experimental realization remain out of reach. Here, we propose and experimentally realize an on-chip photonic system that achieves the non-Abelian braiding of up to five photonic modes. The braiding is realized by controlling the multi-mode geometric-phase matrix in judiciously designed photonic waveguide arrays. The quintessential effect of braiding–sequence-dependent swapping of photon dwell sites is observed in both classical-light and single-photon experiments. Our photonic chips are a versatile and expandable platform for studying non-Abelian physics, and we expect the results to motivate next-gen non-Abelian photonic devices.

	\end{abstract}
	\maketitle
		
	\section{Introduction}
	The existence of non-Abelian anyons in two-dimensional condensed matter systems has been attracting growing attention due to their remarkable features and potential quantum-mechanical applications ~\cite{lei77,wil82m,wil82q,nay08}. When non-Abelian anyons are exchanged by braiding them along the world lines, their wavefunction exchange behaviors are described by a unitary matrix intrinsically different from that of exchanging fermions or bosons~\cite{wil90,fre89}. Since matrix operations are generally non-commutative, the braiding results of non-Abelian anyons are dependent on the order of the braiding operations, which are predicted to be useful for intriguing applications~\cite{nay08}. Although non-Abelian braiding has been dominantly investigated in condensed matter systems~\cite{lut18,wan18,hua21,wil13,bar20,nak20}, its underlying mechanism can be related to the multi-mode geometric-phase effect in classical systems~\cite{zu14,che22}, indicating its universality and compatibility with photonics. Since the very first use of the geometric-phase effect in controlling light polarization in Pancharatnam’s study~\cite{pan56}, various applications have successfully leveraged the single-mode Pancharatnam-Berry phase for light manipulations, such as the light steering with metasurfaces~\cite{zhu20}. However, the exploration of the physical consequence and useful effects of multi-mode geometric phase matrix remains elusive until now. We show here that such effects can be demonstrated by introducing the concept of non-Abelian braiding into photonics.
	
	Recently, a variety of novel non-Abelian phenomena in photonic systems have been predicted and realized~\cite{dal11,oza19,umu13,iad16,dut18,bor19,kre19,che19,yan19,guo21,xu16,xu18,ma16,yan20,bro21,wan21}, including the synthesis of non-Abelian gauge fields~\cite{che19,yan19}, the observation of non-Abelian topological charges~\cite{guo21}, and the simulation of Majorana zero modes in bulk optical systems~\cite{xu16,xu18}. A recent study reports the realization of braiding in the dynamic evolution of one topological mode in a photonic lattice~\cite{noh20}, which produced a phase factor of Abelian nature. However, state permutations are still not directly observed. One of the main challenges lies in the fact that non-Abelian braiding must operate on at least three degenerate states. Therefore, a systematic approach and a versatile, reliable photonic non-Abelian system are highly desirable for exploring the braiding induced multi-mode geometric-phase effect and inspiring novel applications for photon and light manipulations.
	
	In this work, we experimentally realize the non-Abelian braiding of multiple photonic modes on photonic chips. The system is comprised of evanescently coupled photonic waveguides, wherein the evolution of photons follows a Schrödinger-like paraxial equation~\cite{rec13,kla19}. Our scheme leverages chiral symmetry to ensure the degeneracy of multiple zero modes, and drives them in simultaneous adiabatic evolution that induces a unitary geometric-phase matrix that swaps of photon dwell sites. On-chip non-Abelian braiding of up to five modes is observed with both classical light and single photons. We further show that our scheme can be straightforwardly expanded to realize the braiding of even more modes, making the system a versatile platform for studying non-Abelian physics as well as inspiring applications of on-chip non-Abelian photonic devices.
		\begin{figure*}[th]
		\centering
		\includegraphics[width=12cm]{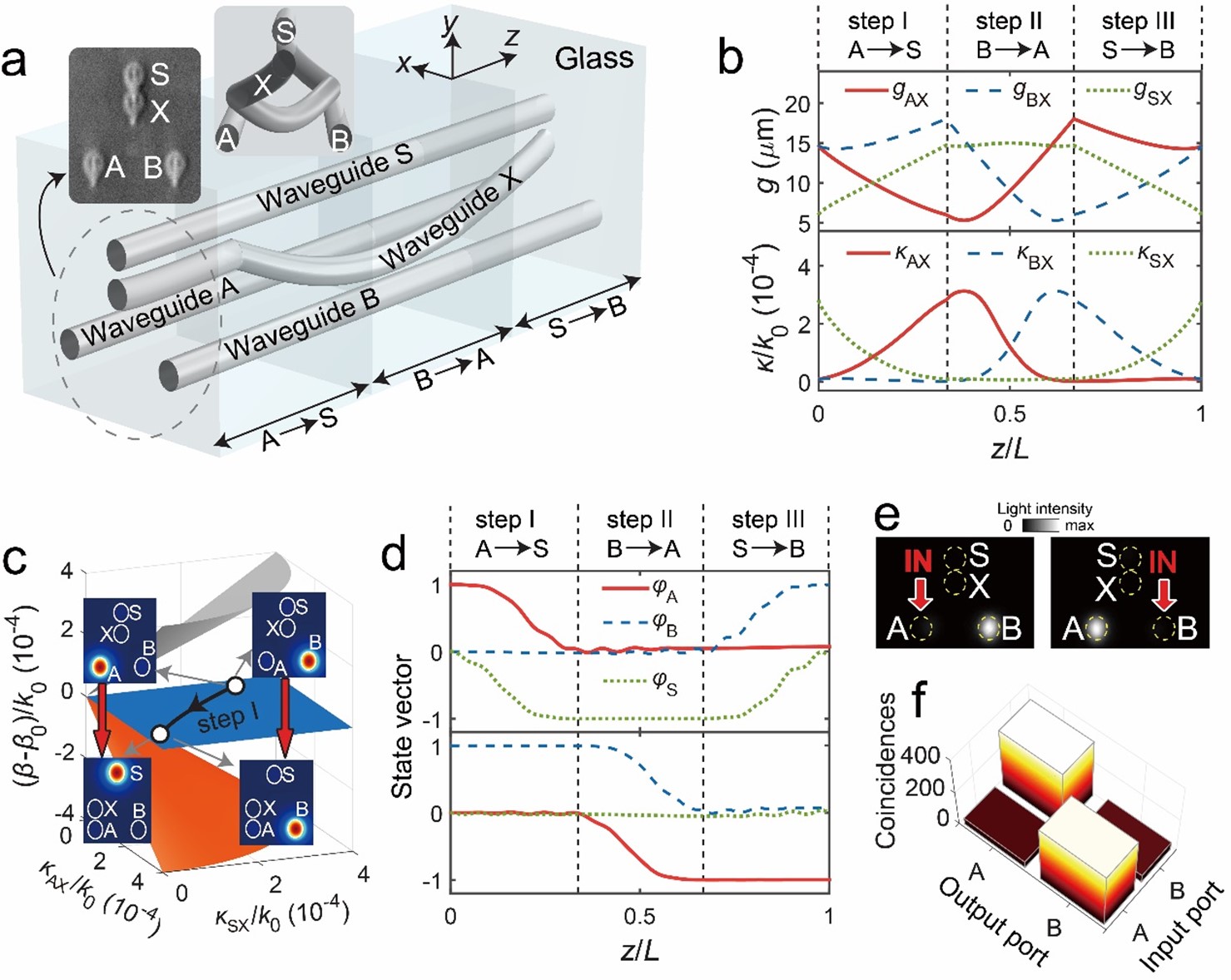}
		\caption{\label{f1}\textbf{Two-mode braiding in photonic waveguides.} \textbf{a,} A schematic diagram of the braiding structure consisting of four waveguides. The inset shows a photograph of the cross-section of the fabricated waveguides. \textbf{b,} The modulation profiles of the waveguide separations (upper) and the corresponding coupling coefficients (lower). \textbf{c,} The eigenvalues of the system as functions of $\kappa_{AX}$ and $\kappa_{SX}$ with $\kappa_{BX}=0$. The blue sheet represents two-fold degenerate states, on which the trajectory depicts the evolution of step I. The corresponding mode exchanges are shown in the insets. \textbf{d,} The calculated state vectors along the braiding direction by setting $\beta_0=0$, for injection at waveguide A (upper) and B (lower). \textbf{e,} The measured light diffraction patterns at the output for injection at waveguide A (left) and B (right). \textbf{f,} The measured coincidences per second at each output waveguide for single-photon injection at different waveguides. }
	\end{figure*}
	\section{Results and discussion}
	\subsection{Two-mode photonic braiding in coupled photonic waveguides}
	We begin our discussion with a two-mode braiding. Figure ~\ref{f1}a illustrates a photonic chip housing the braiding structure consisting of three straight waveguides (A, B, and S) and a curved waveguide X. The straight waveguides have a length of $ L=\ 50\ mm$. These waveguides were fabricated inside boroaluminosilicate glass using femtosecond laser direct writing techniques ~\cite{dav96,yu21} which can induce a refractive index contrast of $\sim 2.5\times 10^{-3}$ between the waveguide and the background. The cross-section of each waveguide measures $\sim6.9\ \mu m\times 5.3\ \mu m$, which meets the single-mode condition for photons polarized in the y-axis at a wavelength of $\sim 810\ nm$. Figure~\ref{f1}b shows the designed center-to-center gap distance $g_{iX}\ \left(i=A,B,S\right)$ between waveguide $i$ and waveguide X and the corresponding coupling coefficient $\kappa_{iX}$. The waveguides A, B, and S are sufficiently separated so that the direct coupling among them is negligible. When the modulations on $\kappa_{iX}$ are sufficiently slow, the dynamics of photon propagation in the waveguides follow a Schrodinger-like equation $\left.H(z)|\psi\left(z\right)\right\rangle={-i\partial}_z\left.|\psi\left(z\right)\right\rangle$, where the Hamiltonian reads
	\begin{equation}\label{e1}
		H\left ( z \right ) =\begin{bmatrix}
			\beta_{X} &\kappa_{AX}\left ( z \right )   & \kappa_{BX}\left ( z \right ) & \kappa_{SX}\left ( z \right )\\
			\kappa_{AX}\left ( z \right )&\beta_{A}  & 0 & 0 \\
			\kappa_{BX}\left ( z \right ) & 0 & \beta_{B} & 0 & \\
			\kappa_{SX}\left ( z \right ) & 0 & 0 & \beta_{S} &
		\end{bmatrix}
		\end{equation}
	In our waveguide system, $\beta_{X,A,B,S}=\beta_0$ which is the waveguide’s propagation constant, and $\left.\left|\psi\left(z\right)\right.\right\rangle=\left[\varphi_X(z),\varphi_A(z),\varphi_B(z),\varphi_S(z)\right]^T$ is the state vector.

	The two-mode braiding is carried out in three steps as indicated in Fig.~\ref{f1}a,b. In step I, $\kappa_{SX} (\kappa_{AX})$ smoothly decreases (increases) from its maximum (zero) to zero (maximum), while $\kappa_{BX}$ is kept at zero. Figure~\ref{f1}c plots the eigenvalues of Eq.(\ref{e1}) as functions of $\kappa_{SX}$ and $\kappa_{AX}$, with $\kappa_{BX}=0$. Two out of the four eigenvalues are independent of the changes in $\kappa_{iX}$ and are always degenerate at a constant eigenvalue $\beta_0$ (see the blue sheet). The two-fold degeneracy is protected by the chiral symmetry $\Gamma ^{-1}H\Gamma=-H$ with\ $\Gamma=\left[\begin{matrix}-1& 0\\0& I_3\end{matrix}\right]$  
	where $I_3$ is a $3\times3$ identity, when we set $\beta_0=0$. We prepare a single-site injection at waveguide A, i.e., $\left.\left|\psi\left(0\right)\right.\right\rangle=\left[0,1,0,0\right]^T$(upper-left inset of Fig.~\ref{f1}c). The adiabatic evolution of $\left.\left|\psi\left(z\right)\right.\right\rangle$ follows the trajectory on the blue sheet and becomes $\left.\left|\psi\left(\frac{1}{3}L\right)\right.\right\rangle=\left[0,0,0,-1\right]^T$, which occupies waveguide S (lower-left inset of Fig. 1c) and picks up a geometric phase of $\pi$. On the other hand, a different injection at waveguide B $\left.\left|\psi\left(0\right)\right.\right\rangle=\left[0,0,1,0\right]^T\rightarrow\left.\left|\psi\left(\frac{1}{3}L\right)\right.\right\rangle=\left[0,0,1,0\right]^T$ remains unchanged (right insets in Fig.~\ref{f1}c). The dynamical phases accumulated for both injections are $\beta_{0}L/3$. Therefore, the total phases accumulated in the above two processes differ by $\pi$ which is the geometric phase.

	Steps II and III can be understood similarly. In step II, the state dwelling in waveguide B is relocated to A and acquires a geometric phase $\pi$, i.e., $\left.\left|\psi\left(\frac{1}{3}L\right)\right.\right\rangle=\left[0,0,1,0\right]^T\rightarrow\left.\left|\psi\left(\frac{2}{3}L\right)\right.\right\rangle=\left[0,-1,0,0\right]^T$. In step III, and the state occupying waveguide S with $\left.\left|\psi\left(\frac{2}{3}L\right)\right.\right\rangle=\left[0,0,0,-1\right]^T$ transfers to waveguide B state $\left.\left|\psi\left(L\right)\right.\right\rangle=\left[0,0,1,0\right]^T$ and also obtains the $\pi$ phase. The mode switching behaviors can be verified by the calculated state vectors in the braiding process, as plotted in Fig. 1d. To summarize, the net outcome of the evolution is the swapping of the states in waveguides A and B, i.e., $\left.\left|\psi\left(0\right)\right.\right\rangle=\left[0,1,0,0\right]^T\rightarrow\left.\left|\psi\left(L\right)\right.\right\rangle=\left[0,0,1,0\right]^T$ and $\left.\left|\psi\left(0\right)\right.\right\rangle=\left[0,0,1,0\right]^T\rightarrow\left.\left|\psi\left(L\right)\right.\right\rangle=\left[0,-1,0,0\right]^T$, and the output states differ by a geometric phase of $\pi$ (the accumulated dynamical phases are both $\beta_{0}L$ which is omitted in the expression). This result can be captured by a unitary matrix $Y=\left[\begin{matrix}0&-1\\1&0\\\end{matrix}\right]$, which is a $U\left(2\right)$ operation also known as the $Y$-gate in quantum logics~\cite{bar95}.
	
	We show experimental results to verify the design. Injection at waveguide A and B is performed with a laser at 808 nm (CNI, MDL-III-808L). The light diffraction patterns at the output facet were recorded using a CCD (XG500, XWJG). The photographs are shown in Fig.~\ref{f1}e, where the swapping of light-dwelling sites is clearly seen. The braiding was also verified in the quantum-mechanical limit by single-photon injections, where indistinguishable pairs of photons at 810 nm were generated using a quantum setup. One photon was injected into the braiding structure via waveguide A or B, while the other one propagated in a single-mode reference optical fiber. We used two avalanche photodetectors to respectively collect the single photons at the output (i.e., waveguide A or B) and the reference fiber. Coincidence measurements were then performed using the collected data. In the results displayed in Fig.~\ref{f1}f, the output waveguide exhibiting the dominated coincidence is always different from the input one – clear evidence of the two-mode braiding described by the $Y$-gate.
	\subsection{Measurement of the geometric phase}
\begin{figure}[t]
		\centering
		\includegraphics[width=8.4cm]{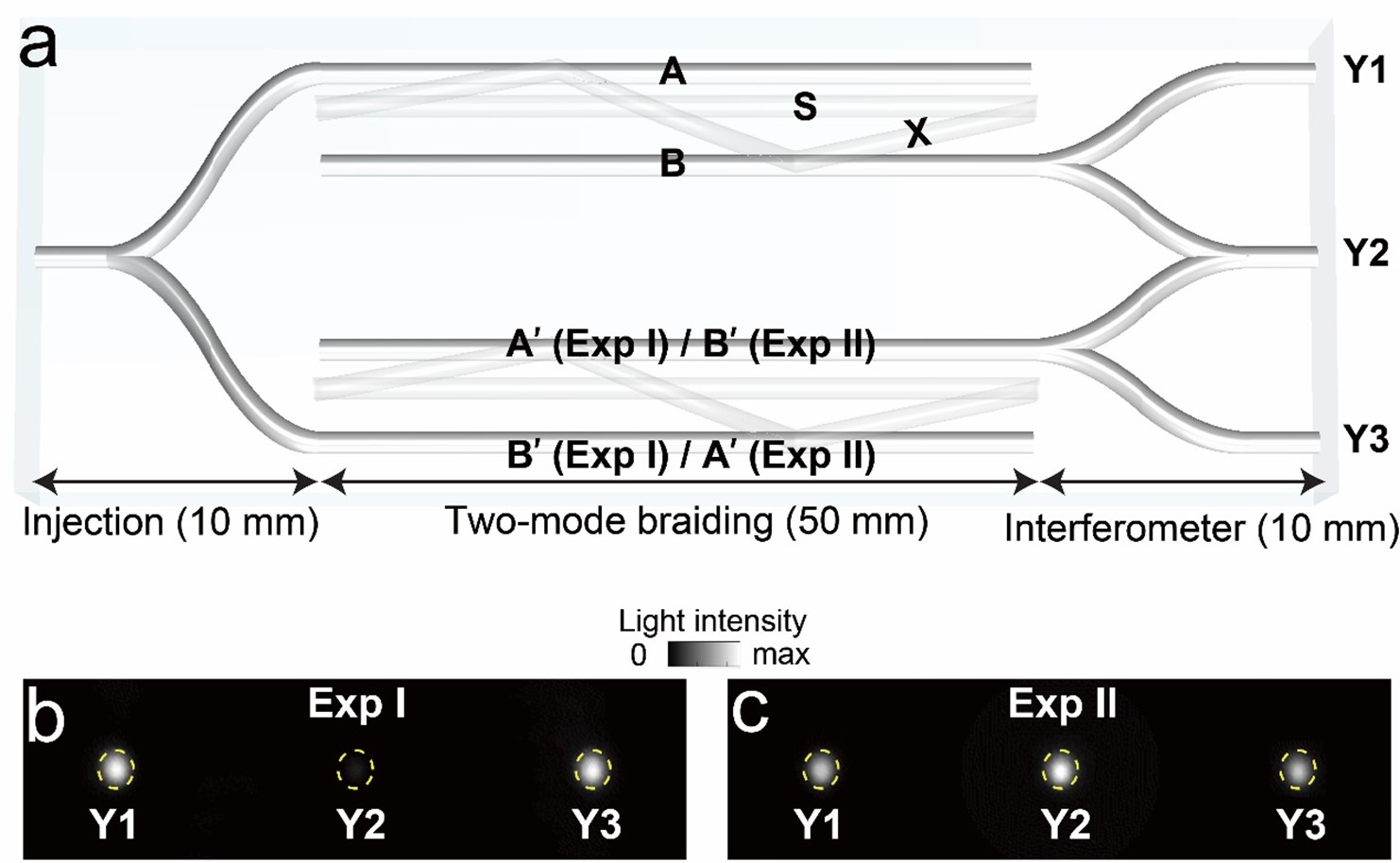}
		\caption{\label{f2}\textbf{Measurement of the geometric phase in two-mode braiding.} \textbf{a,} A schematic diagram of interference experiments. Two four-waveguide clusters produce the same braiding with different or identical injections and the interference of their output states at Y2 port reveals the geometric phase. \textbf{b,c,} The measured light diffraction patterns at the output for injections at $\mathrm{A}$ and $\mathrm{B}^{\prime}$ (Exp I) \textbf{(b)} and at $\mathrm{A}$ and $\mathrm{A}^{\prime}$ (Exp II) \textbf{(c)}. In \textbf{(b)}, Y2 has almost no power output, indicating a destructive interference. In \textbf{(c)}, Y2 lights up because the two arms interfere constructively.}
	\end{figure}

	To measure the geometric phase that is a key characteristic of the two-mode braiding, we have designed an interference experiment, as illustrated in Fig.~\ref{f2}a. The structure consists of three stages. In the injection stage, injected photons are equally split into two identical waveguides. Two separate braidings are then carried out in the second stage which contains two copies of braiding structures identical to the one in Fig.~\ref{f1}a. Two experiments are performed with different configurations. In experiment I (II), injection of the lower braiding structure is via waveguide $\mathrm{B}^{\prime} (\mathrm{A}^{\prime })$. The upper braiding structure is injected at waveguide A for both experiments. After the braiding, in stage III, the two output waveguides are equally split into four arms, two of which are merged again so that photons from the upper and lower clusters can interfere. The three terminal ports are labelled Y1, Y2, and Y3.
	
	The results of the two experiments with light are shown in Fig.~\ref{f2}b,c. Strong light intensities are seen at ports Y1 and Y3 for both cases, which indicates the successful braiding induced mode switching. However, discrepancies are seen at Y2. In experiment I (injections at $\mathrm{A}$ and $\mathrm{B}^{\prime}$), the image at Y2 is dark (Fig.~\ref{f2}b), which suggests a destructive interference of the light after braiding. Because the light propagating through the upper and lower braiding structures accumulates the same dynamical phase, this result indicates a phase difference of $\pi$, which can only be the consequence of a geometric phase. In experiment II for comparison (injections at $\mathrm{A}$ and $\mathrm{A}^{\prime}$), the port Y2 lights up (Fig.~\ref{f2}c), which suggests a constructive interference due to their same phase accumulation. These experimental results are strong evidence of the $\pi$ geometric phase difference induced by the two-mode braiding.
	\begin{figure*}[ht]
		\centering
		\includegraphics[width=14cm]{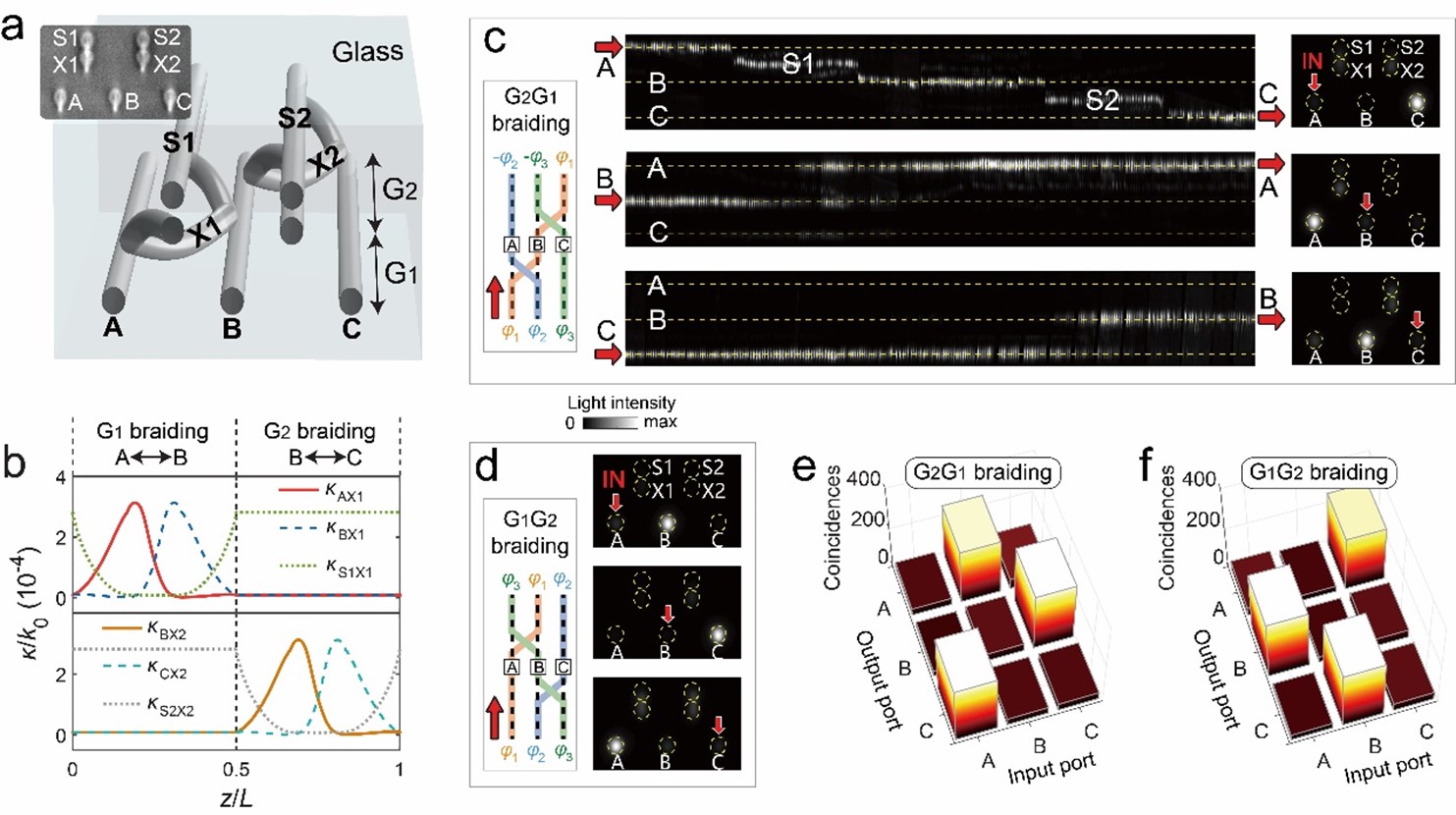}
		\caption{\label{f3}\textbf{Non-Abelian braiding of three modes.} \textbf{a,} A schematic diagram of a $G_2G_1$ braiding configuration. \textbf{b,} The modulation profiles of the coupling coefficients. \textbf{c,} Experimental results of the $G_2G_1$ braiding (left: braid diagram where the black dashed lines represent the waveguides and the colored lines mark the braiding path of wavefunctions), including the measured light diffraction patterns in the evolution process (middle) and the output light intensity distributions (right). The red arrows mark the injection and output waveguides. \textbf{d,} Measured light diffraction patterns at the output facet of the $G_1G_2$ braiding. \textbf{e,f,} The measured coincidences per second at each output waveguide with single-photon injections for $G_2G_1$ \textbf{(e)} and $G_1G_2$\textbf{(f)}. The distinct outcomes of the two configurations are clear evidence of the non-Abelian nature of the braiding.}
	\end{figure*}
	\subsection{Three-mode non-Abelian braiding}
	The two-mode braiding demonstrates the effectiveness of our photonic platform, which we now expand for three modes. The three-mode braid group has two generating operations $G_1:\left.\left|\psi^\prime\right.\right\rangle=[\varphi_{1},\varphi_{2},\varphi_{3}]^T\rightarrow[-\varphi_{2},\varphi_{1},\varphi_{3}]^T,$ $G_2:\left.\left|\psi^\prime\right.\right\rangle=[\varphi_{1},\varphi_{2},\varphi_{3}]^T\rightarrow[,-\varphi_{3},\varphi_{2}]^T$, where $\left.\left|\psi^\prime\right.\right\rangle$ is a truncated state vector with its elements successively denoting the wavefunction in waveguide A, B and C, respectively. Unlike the permutation of two modes which has only one possibility, permutations of three or more modes are non-Abelian in character, i.e., $G_2G_1\neq G_1G_2$, which we will next demonstrate. Figure~\ref{f3}a illustrates the schematic diagram for the three-mode braiding structure. Here, the system has seven waveguides. Waveguides A, B, C are sufficiently far apart so that they can only couple via waveguides X1 and X2. The system’s Hamiltonian thus has a similar structure as Eq.(\ref{e1}) but sustains three degenerate zero modes that form the braiding subspace. The waveguide array has a length of 80 mm and is divided into two sections, with the fitted coupling coefficients shown in Fig.~\ref{f3}b. The first section swaps the modes in waveguides A and B, which executes $G_1$. The second section exchanges modes in B and C, so that the net result is $G_2G_1:[\varphi_{1},\varphi_{2},\varphi_{3}]^T\rightarrow[-\varphi_{2},-\varphi_{3},\varphi_{1}]^T$ (see the braiding diagram in the left panel of Fig.~\ref{f3}c). The experimental results of $G_2G_1$ are summarized in Fig.~\ref{f3}c. We employ a double exposure-assisted scattering technique, in which point-like scatterers were fabricated inside all the waveguides so that the light passages can be captured by a camera (Zyla 5.5 sCMOS, Andor) as shown in the middle panels. We find that when the injection is at waveguide A, the light successively propagates to the waveguide S1, B, S2, and finally outputs at C. In contrast, when injected at waveguide B (C), the output is at A (B). The output patterns are shown in Fig.~\ref{f3}c (right panels), which are clearly the intended three-mode permutation described by $G_2G_1$.
	
	We fabricated another system with two sections arranged in the opposite order so that $G_1G_2:[\varphi_{1},\varphi_{2},\varphi_{3}]^T\rightarrow[\varphi_{3},\varphi_{1},\varphi_{2}]^T$ is executed (see the braiding diagram in the inset of Fig.~\ref{f3}d). The experimental results shown in Fig.~\ref{f3}d demonstrate the intended outcome. Comparing Fig.~\ref{f3}c and d, it is evident that $G_2G_1\neq G_1G_2$, which unambiguously demonstrates the non-Abelian nature of the three-mode braiding.
	
	The non-Abelian braiding is also successfully realized in single-photon experiments using the same photonic chips. The measured coincidences confirm that single photons also follow non-Abelian braiding, as shown in Fig.~\ref{f3}e,f for $G_2G_1$ and $G_1G_2$, respectively.
	
	Another property of three-mode braiding is the equivalence of $G_2G_1G_2$ and $G_1G_2G_1$~\cite{lut18}, both of which induce $[\varphi_{1},\varphi_{2},\varphi_{3}]^T\rightarrow[\varphi_{3},-\varphi_{2},\varphi_{1}]^T$. These braiding operations can be used to design a quantum $X$-gate (by discarding one state after the operation), which is confirmed in separate experiments.
	\begin{figure*}
		\centering
		\includegraphics[width=11cm]{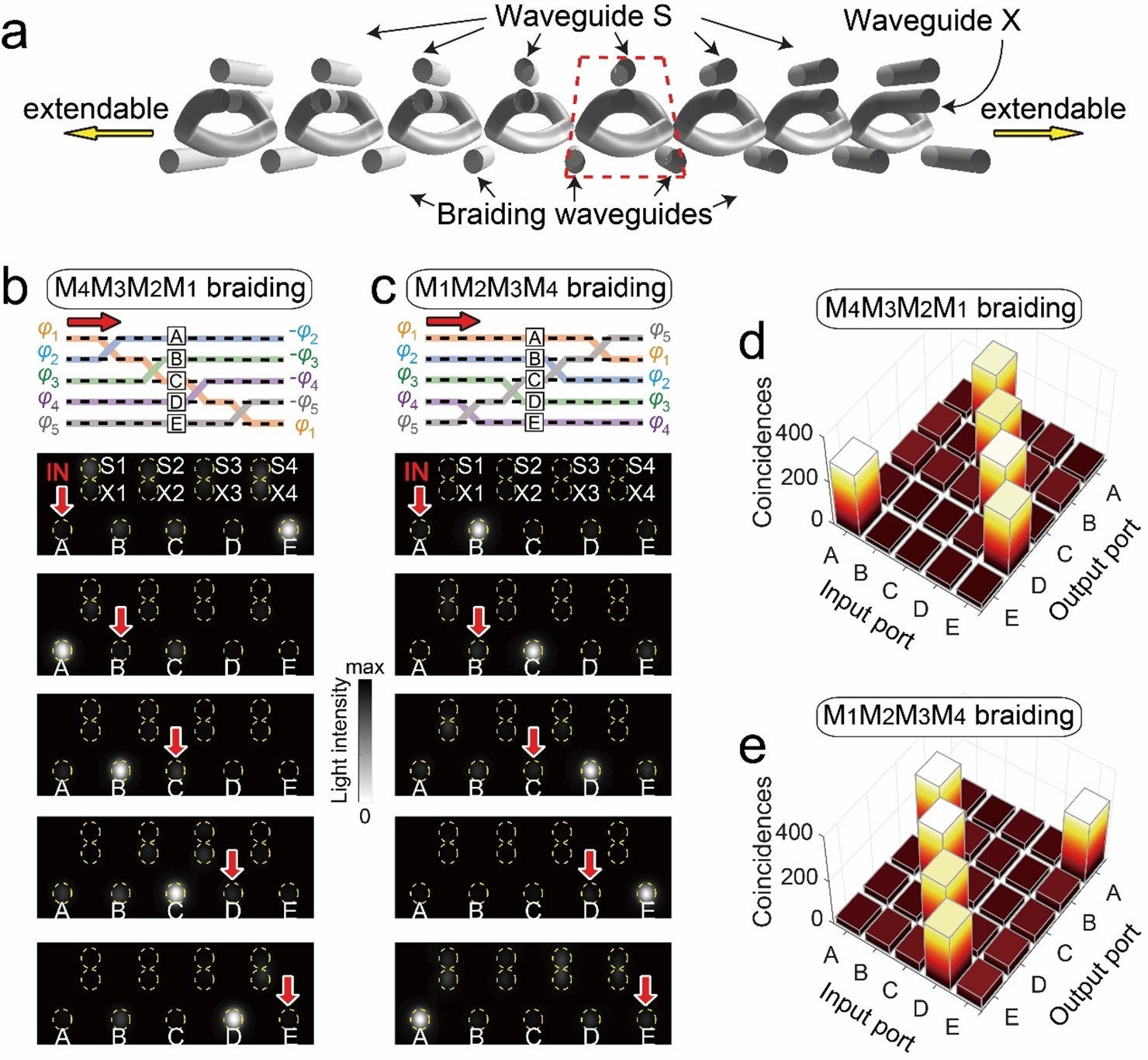}
		\caption{\label{f4}\textbf{Multi-mode braiding.}\textbf{ a,} The expandability of the braiding design.  \textbf{b,c,} Two examples of five-mode braiding (upper: braid diagram) and their experimental realization with light (lower). \textbf{d,e,} Results of single-photon five-mode braiding experiments.}
	\end{figure*}
	\subsection{Expandability of the photonic chips and multi-mode braiding}
	Comparing the structures of two-mode and three-mode braiding, it becomes clear that the photonic platform can be straightforwardly expanded to realize the braiding of an arbitrary number of modes, as shown in Fig.~\ref{f4}a. As a proof of concept, we present a five-mode braiding design. The targeted braid diagrams are shown in Fig.~\ref{f4}b,c (upper), which depict two operation sequences: $M_4M_3M_2M_1: [\varphi_{1},\varphi_{2},\varphi_{3},\varphi_{4},\varphi_{5}]^T\rightarrow[-\varphi_{2},-\varphi_{3},-\varphi_{4},-\varphi_{5},\varphi_{1}]^T$ and $M_1M_2M_3M_4: [\varphi_{1},\varphi_{2},\varphi_{3},\varphi_{4},\varphi_{5}]^T\rightarrow[\varphi_{5},\varphi_{1},\varphi_{2},\varphi_{3},\varphi_{4}]^T$. The measured diffraction patterns using lasers are given in Fig.~\ref{f4}b,c, and the measured coincidences using single photons are summarized in Fig.~\ref{f4}d,e. The observed permutation outcomes align well with theoretical predictions.
	
	We further remark that within each building block (red box in Fig.~\ref{f4}a), only one waveguide needs to be bent to achieve the modulation of three coupling coefficients. As a result, the top and bottom rows are all identical straight waveguides that can be prefabricated, which makes the design of subsequent braiding quite flexible. With these characteristics and the expandability as demonstrated, our scheme becomes a versatile and convenient on-chip platform for realizing more complex non-Abelian operations.
	\section{Conclusion}
	To conclude, we have realized non-Abelian braiding for both classical light and single photons in a photonic on-chip platform. The key characteristics of non-Abelian braiding – permutations of degenerate states and the order-dependent braiding outcomes, are both definitively observed. We remark that our scheme for photonic non-Abelian braiding is a purely geometric-phase effect on multiple degenerate modes protected by chiral symmetry. As a result, the braiding is robust to perturbations such as those on the evolution path and coupling coefficients. How to incorporate topological protection to make the braiding operations even more robust is a valuable future goal. The proposed versatile photonic platform is expected to reveal more non-Abelian physics related to the multi-mode Berry phase and its capability in generating a variety of unitary matrices may lead to a new generation of non-Abelian photonic devices for unprecedented light and photon manipulations.

\bibliography{reference}	
	
\end{document}